\documentclass[submission,copyright,creativecommons]{eptcs}
\providecommand{\event}{ICLP 2023} 

\usepackage{iftex}

\usepackage[table,xcdraw]{xcolor}
\usepackage{graphicx}    

\usepackage{caption}     

\usepackage{listings}
\lstdefinestyle{prologStyle}{
    language=Prolog,
    basicstyle=\small\ttfamily,
    commentstyle=\color{gray},
    keywordstyle=\color{blue},
    morekeywords={:-},
    frame=single,
    numbers=left,
    numberstyle=\tiny,
    stepnumber=1,
    numbersep=5pt,
    breaklines=true,
    tabsize=4,
    showspaces=false,
    showstringspaces=false,
    literate={:-}{{\textcolor{blue}{:-}}}2
}

\ifpdf
  \usepackage{underscore}         
  \usepackage[T1]{fontenc}        
\else
  \usepackage{breakurl}           
\fi

\title{Inferring Compensatory Kinase Networks in Yeast using Prolog}
\author{George A. Elder
\institute{Queen Mary University of London}
\institute{School of Biological and Behavioural Sciences \\
QMUL \thanks{Research partially funded by
      Medical Research Council MRC}\\
London, United Kingdom}
\email{g.a.elder@qmul.ac.uk}
\and
Conrad Bessant
\institute{Queen Mary University of London}
\institute{School of Biological and Behavioural Sciences \\
QMUL \\ London, United Kingdom}
\email{c.bessant@qmul.ac.uk}
}
\def\titlerunning{Inferring Compensatory Kinase Networks in Yeast using Prolog}
\def\authorrunning{George A. Elder \& Conrad Bessant}
\begin{document}
\maketitle

\begin{abstract}
Signalling pathways are conserved across different species, therefore making yeast a model organism to study these via disruption of kinase activity. Yeast has 159 genes that encode protein kinases and phosphatases, and 136 of these have counterparts in humans. Therefore any insight in this model organism could potentially offer indications of mechanisms of action in the human kinome. This study utilises a Prolog-based approach, data from a yeast kinase deletions strains study and publicly available kinase-protein associations. Prolog, a programming language that is well-suited for symbolic reasoning is used to reason over the data and infer compensatory kinase networks. This approach is based on the idea that when a kinase is knocked out, other kinases may compensate for this loss of activity. Background knowledge on kinases targeting proteins is used to guide the analysis. This knowledge is used to infer the potential compensatory interactions between kinases based on the changes in phosphorylation observed in the phosphoproteomics data from the yeast study. The results demonstrate the effectiveness of the Prolog-based approach in analysing complex cell signalling mechanisms in yeast. The inferred compensatory kinase networks provide new insights into the regulation of cell signalling in yeast and may aid in the identification of potential therapeutic targets for modulating signalling pathways in yeast and other organisms.
\end{abstract}

\section{Introduction}

As with any organism, so with \textit{Saccharomyces cerevisiae}, protein phosphorylation plays a crucial role in various signalling pathways that control cellular activities such as gene expression, protein synthesis, DNA replication, RNA processing, cell cycle regulation, metabolism, transport of vesicles and other organelles, response to environmental stress and nutrients, and cell differentiation. The specific actions of kinases and substrates are largely determined by protein interactions, which are effected via phosphorylation \cite{Ptacek2005}.

It is also known that many of these signalling pathways are conserved across different species, so yeast can be used as a model organism to study phosphorylation networks. Yeast has 159 genes that encode protein kinases and phosphatases, and 136 of these have counterparts in humans \cite{Manning2002}. Therefore any insight in this model organism could potentially offer indications of mechanisms of action in the human kinome. 

Ours is not the first application of symbolic AI in the study of yeast and its pathways more specifically. A major inspiration for applying techniques that fall under the symbolic framework came from the extensive work done by Prof R.King. Particularly, work of the type found in this study by A.Clare and R.King \cite{dataminingClareKing} showed that it is possible to build a knowledge base via the use of a declarative logic programming language. The seminal work on Robot Scientist Project also offered insight in how logic programming (in this case inductive) can be used to model the metabolism of yeast and how deletion in associated genes can affect their respective processes \cite{ROBOKING}. Further use of a Prolog based approach can be found in \cite{Whelan2008} where a Flux Balance Analysis model was combined with genome related information from the KEGG database to predict the growth patterns of yeast. More recently, network reconstruction as well as new structure hypotheses were made using ProbLog, with the aim of providing new insights to experimental biologists \cite{Goncalves2014}.

However, all of the above approaches have focused on the use of gene-based background knowledge and experimentally derived facts to reach their conclusions. These lack the granularity of a proteomics and even more so a phosphoproteomics based approach.

\section{Methods}

\subsection{Toy Model development}

Initially a Toy Model version of our logic program was developed, in order to test the feasibility and applicability of our proposed methodology. The perturbations modelled were gene deletions and the question thus became whether silencing (or deletion) of a given kinase has actually worked. Within the context of the study chosen (to be described in detail in subsequent section), specific gene deletions were confirmed by either proteomics or PCR assays. Therefore and due to the nature of gene deletions in yeast it was unnecessary to identify off targets or known targets of a given perturbation.

\begin{figure}[ht!]
\centering
\includegraphics[width=.9\textwidth]{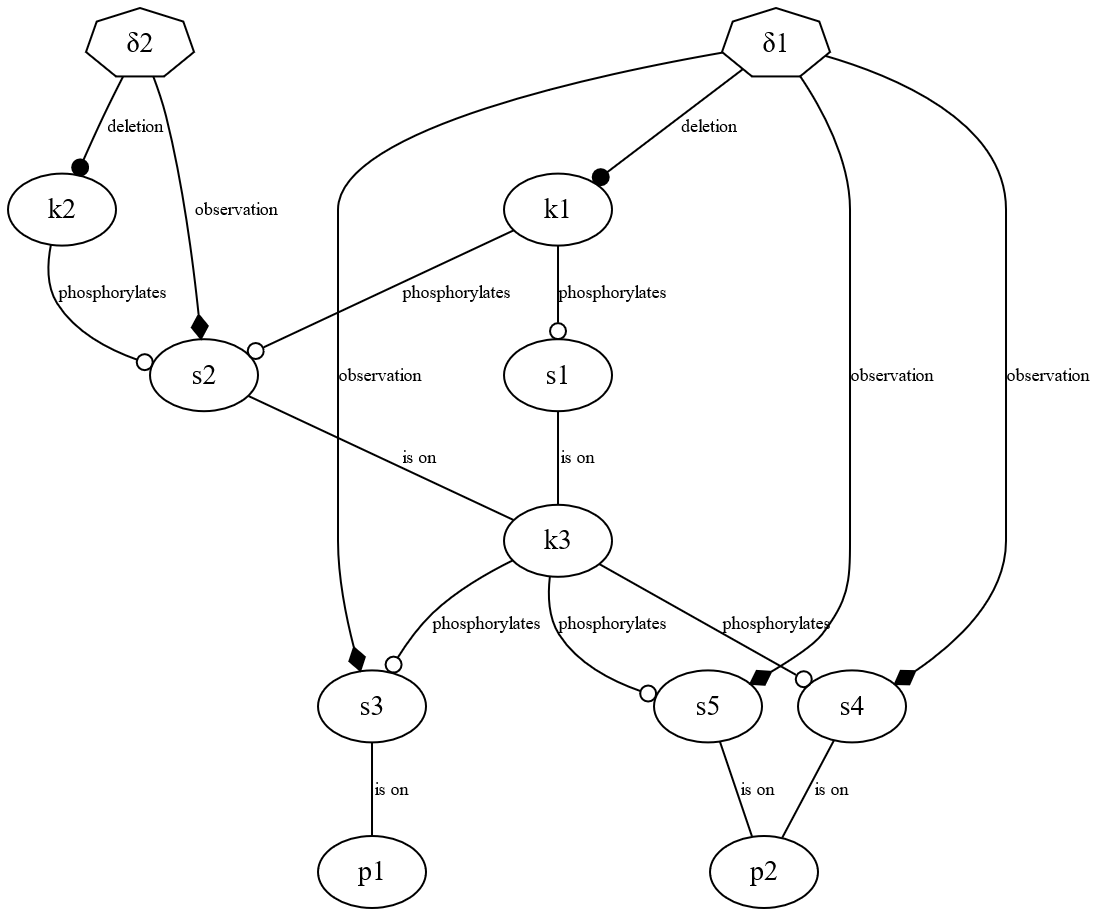}
\caption[Yeast Toy Model] {A Toy Model illustrates the minimal number of interactions needed to understand a biological scenario involving kinases (k), proteins (p), and their phosphosites (s) in response to an initial perturbation (\textbf{$\delta$}). In our model, the perturbation refers to a gene encoding a specific kinase being deleted. Our background information in this implementation has levels of granularity, in this case the kinase targeting a phosphosite on a protein version is depicted.}
\label{fig:yeastToyModel2}
\end{figure}

In Figure \ref{fig:yeastToyModel2} the overall structure of our yeast Toy Model can be seen. Edges represent either perturbations (deletions in this case), observational \textit{facts}, presence of a phosphosite on a kinase and kinase phosphosite associations. Nodes represent either phosphosites, perturbations, kinases or other proteins. 


To determine whether a kinase has been successfully silenced, development of the logic program began with the first iteration of \textit{rules}. When queried, it is satisfied in the case where a kinase \textbf{K} (1), for which a known phosphosite \textbf{S} target on a protein \textbf{P} is found (2). Lastly, (3) checks whether a corresponding mass spectrometry \textit{fact} following deletion of \textbf{K} can be matched. If all these conditions are met then the interpreter responds with {\fontfamily{qcr}\selectfont true} thus confirming that the $\delta$gene perturbation encoding kinase \textbf{K} has been successful. \\

\begin{lstlisting}[style=prologStyle]
doesSilofKwork(deletionOF((kinase(K))) :-
    phosphorylates(kinase(K), phosphosite(S), protein(P)),
    perturbs(deletionOF((kinase(K))),phosphosite(S), occupancy(down),_).
\end{lstlisting}

In the next \textit{rule}, developed in an iterative manner, a ``uniqueness check'' was added. This establishes whether a phosphosite \textbf{S} is uniquely targeted by a kinase \textbf{K}. This is achieved by establishing a \textit{rule} and checking for its negation as part of this second iteration. In Prolog notation:

\begin{lstlisting}[style=prologStyle]
doesSilofKwork2(deletionOF((kinase(K))) :-
    phosphorylates(kinase(K), phosphosite(S), protein(P)),
    \+ sharedtarget(kinase(K), kinase(_K2), phosphosite(S)),
    perturbs(deletionOF((kinase(K))),phosphosite(S), occupancy(down),_).
\end{lstlisting}

The \textit{rule} {\fontfamily{qcr}\selectfont sharedtarget(kinase(K), kinase(\_K2), phosphosite(S))} includes two goals matching phosphorylation \textit{facts} on shared phosphosites then checking that the two phosphorylating are not the same. \\

\begin{lstlisting}[style=prologStyle]
sharedtarget(kinase(K), kinase(K2), phosphosite(S)) :-
    (phosphorylates(kinase(K), phosphosite(S), kinase(P)),
    phosphorylates(kinase(K2), phosphosite(S), kinase(P)),
    K1 \= K2).
\end{lstlisting}

In the toy model, the responses to these two iterations were as expected. However, there is still a concern that when applied to the larger dataset with more than five perturbation-affected phosphosites, the true responses will be based on only one phosphosite instance. Therefore the addition of a majority check was decided. The final version of the \textit{rule} took the following form:

\clearpage
\begin{lstlisting}[style=prologStyle]
doesSilofKwork3(deletionOF((kinase(K))) :-
    phosphorylates(kinase(K), phosphosite(S), kinase(P),
    \+ sharedtarget(kinase(K), kinase(_K2), phosphosite(S)),
    findall((phosphosite(S),
    perturbs(deletionOF((kinase(K))),phosphosite(S), occupancy(up),_),
    perturbs(deletionOF((kinase(K))),phosphosite(S), occupancy(unaffected),_),
    List1)),
    findall((phosphosite(S),
    perturbs(deletionOF((kinase(K))),phosphosite(S), occupancy(down),_),
    List2)))),
    length(List1, Length1), length(List2, Length2),
    compare( '>', Length2, Length1).
\end{lstlisting}

This \textit{rule} adds the majority check to the previous \textit{rule} version. This creates two lists, one containing all known targets of the kinase that are either ``up'' or ``unaffected'' and another that contains all the ones described as ``down'', then compares the length of the two. The aim of this is to establish whether the majority of the known targets were ``down'' rather than not, therefore confirming the silencing of the kinase.

\subsection{Experimental \textit{facts} and background knowledge sources}

A recently published study that looked at the effect of kinase perturbation at the phosphoproteomic level was chosen as the source of our experimental data. In \cite{Li2019}, a systems-level proteomic and phosphoproteomic analysis was carried out on 110 yeast single-kinase or phosphatase deletion strains under standard growth conditions. The 110 were split in 84 kinases and 26 phosphatases. They employed various methods, including traditional enrichment analysis, $\delta$gene-$\delta$gene correlation networks, and molecular covariance networks in order to analyse the functional relationships between these active proteins. 

In total their chosen deletions, which will be referred to as perturbations in this section, cover 82\% of all possible yeast and phosphatase deletion strains. Through their experimental and mass spectrometry workflow they were able to identify more than 4,600 and 13,000 proteins and phosphosites, respectively. Of interest is the finding that they were able to capture a large part of regulated phosphorylation events as well as 30\% of those being newly captured. This was entirely attributed to their ability to normalise their data with protein abundance. This is also the data that we elected to include in our 'perturbs/n' predicate format. 

Key constituents of our background knowledge base are the associations between kinases and their targets both at the protein and phosphosite levels. For this, two main sources were chosen, namely the Yeast Kinase Interaction Database \cite{Sharifpoor2011} (accessible at \url{http://www.moseslab.csb.utoronto.ca/KID/}) and the Yeast Kinome database \cite{Breitkreutz2010} (accessible at \url{https://thebiogrid.org/project/2}).

The Yeast Kinase Interaction Database (KID) is a resource that contains data on phosphorylation events from various high- and low-throughput experiments. It includes a total of literature-curated low-throughput (6,225) and high-throughput (21,990) interactions, stemming from over 35,000 experiments. It includes 517 high-confidence (or 853 low-confidence) kinase-substrate pairs depending on the cutoffs used for the metric provided in the study. On the other hand the Yeast Kinome database contains 1,844 interactions observed by mass-spectrometry based analysis of protein complexes. It is updated monthly and has been subsumed as part of the BioGRID project which, in turn is a broad interaction database including more than 1.7m individual interactions \cite{Oughtred2021}.

Both these contain kinase (and phosphatase) target associations with the latter (Yeast Kinome db) also containing these at the phosphosite level. As will be shown below however, there was an issue of overlap between our experimental \textit{facts} and background knowledge which limited the predictions that could be made. 

\subsubsection{Yeast perturbation observation \textit{facts} and background knowledge sources}

From the perturbation study chosen, a total of 685,358 \textit{facts} were extracted containing the effect of the aforementioned 110 kinase and phosphatase deletions over 13,258 individual phosphosites. They were extracted from a .csv file containing the log2fold change recorded for each phosphosite between wild type and $\delta$gene strains of yeast (i.e. strains where selected genes have been deleted). For most phosphosites there were two replicates reported. Not all phosphosites had both replicates for all $\delta$gene combinations. Only phosphosites containing two replicates were considered and the mean value between the two was taken for those that did. The code determined whether the average of the log2 fold change values is less than or equal to -0.5, greater than or equal to 0.5, or between those two values. Depending on the value of the average, the code sets the value of a variable called `` change'' to ``down'', ``up'', or ``unaffected'', respectively. \textit{Facts} took the following form: \\ 

{\fontfamily{qcr}\selectfont perturbs('YCK2', 'ENO2\_pT324', 'ENO2', unaffected, 0.0475).}\\

The background knowledge base includes \textit{facts} that describe the relationships between kinases, proteins and phosphosites at both the kinase-protein and kinase-phosphosite level. These were taken from the two databases described above and were transcribed into \textit{facts} in the following format: \\

Protein level information: 
{\fontfamily{qcr}\selectfont knownKtarget('ARK1', 'PAN1').}

Phosphosite level information: 
{\fontfamily{qcr}\selectfont knowntarget('ATG1','ALY1\_pS813').} \\

The first set of \textit{facts} describing kinase - protein level information came from the Kinome Interaction Database and the Yeast KINOME database, while the kinase - phosphosite level information came from the KINOME/bioGRID database alone. \\

All scripts used during data and background knowledge curation and transcription as well as .pl Prolog files can be found at \url{https://github.com/Dudelder/Yeast_Prolog_Compensatory}). Total number of \textit{facts} that make up knowledge base can be seen in Table 1.


\begin{table}
\centering
\begin{tabular}{|l|l|l|}
\hline
Facts List & Source & \begin{tabular}[c]{@{}l@{}}Size\\ (Number of facts)\end{tabular} \\ \hline
\begin{tabular}[c]{@{}l@{}}Dgene-Dgene phosphosite\\ phosphosite observations\end{tabular} & \begin{tabular}[c]{@{}l@{}}Perturbation\\ Study\end{tabular} & 685,358 \\ \hline
Kinase Substrate Associations & \begin{tabular}[c]{@{}l@{}}Yeast Kinome\\ Database / bioGRID\end{tabular} & 7589 \\ \hline
Kinase Protein Associations & \begin{tabular}[c]{@{}l@{}}Yeast Kinome\\ Database / bioGRID\end{tabular} & 1558 \\ \hline
Kinase Protein Associations & \begin{tabular}[c]{@{}l@{}}Kinase Interaction\\ Database (lenient)\end{tabular} & 853 \\ \hline
Kinase Protein Associations & \begin{tabular}[c]{@{}l@{}}Kinase Interaction\\ Database (Strict)\end{tabular} & 517 \\ \hline
 & Total = & 695,875 \\ \hline
\end{tabular}
\caption[Yeast Background Knowledge base]{Table listing the title, sources and associated number of \textit{facts} that make up the Background Knowledge section of the yeast logic model. The total reported is for the strict cut-off of the Kinome Interaction Database. For the lenient cut-off, the total is 696,211. Specific sources referenced \cite{Sharifpoor2011}, \cite{Li2019}, \cite{Breitkreutz2010}.}
\end{table}

\subsubsection{Overlap between experimental observations and background knowledge}

\begin{figure}[ht]
    \centering
\includegraphics[width=.45\textwidth]{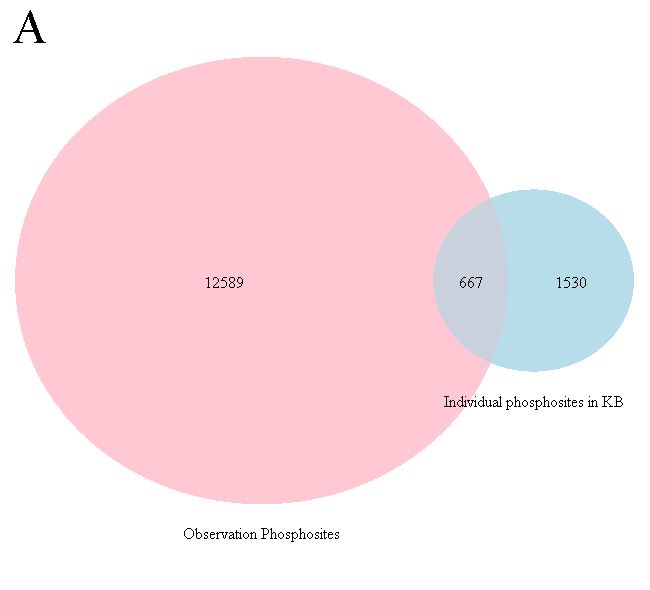}
\includegraphics[width=.45\textwidth]{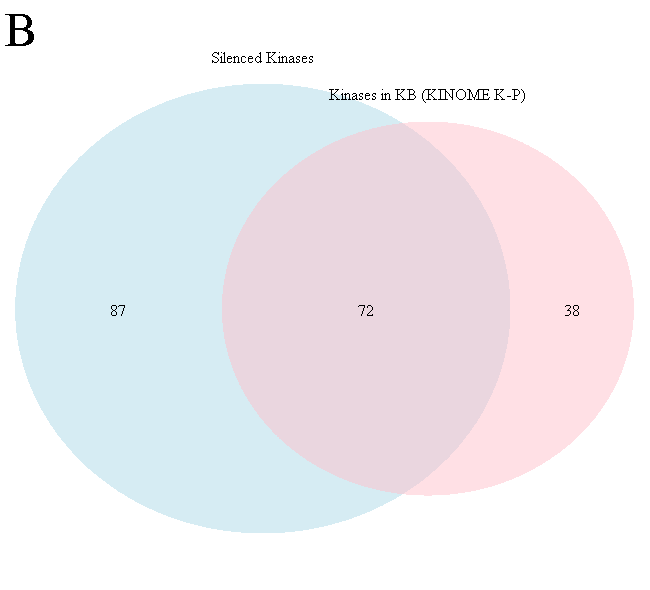}
\includegraphics[width=.6\textwidth]{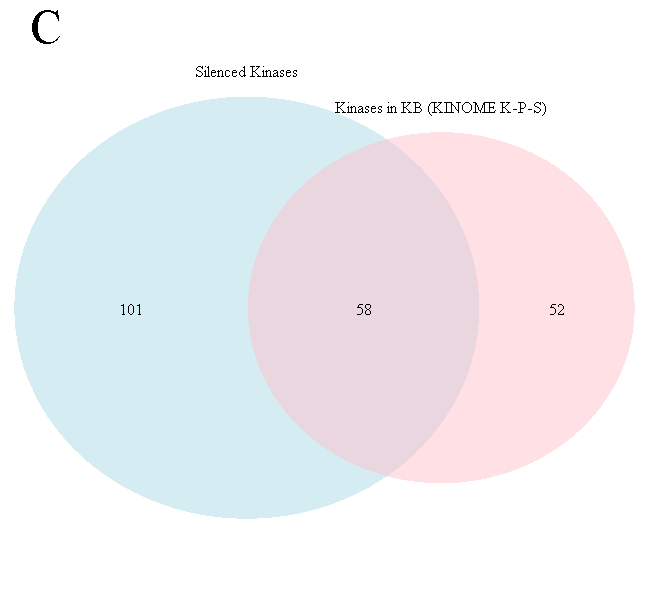}
\captionsetup{font=small}
    \caption[Background Knowledge and Observational \textit{facts} Overlap for Kinome DB] {Venn Diagrams depicting the overlap between our experimental data and background knowledge sourced from Kinome Database / bioGRID \cite{Breitkreutz2010}. In A) total number of phosphosites is 13,256. B: Overlap between kinases in KinomeDB/bioGRID and deleted kinases at the kinase-protein level of association. C: Overlap between kinases in KinomeDB/bioGRID and deleted kinases at the kinase-phosphosite level of association.}
    \label{fig:overlap}
\end{figure}

There is an overlap between the observational \textit{facts} at the kinase-phosphosite level and the phosphosites in the background knowledge base. Specifically, out of the total number of individual phosphosites, only 667 are found in the kinase-phosphosite associations knowledge base (Figure \ref{fig:overlap} A), accounting for approximately 5$\%$ of the total.. At the protein level things improve slightly depending on the data source and cutoff used (for Kinase Interaction Database). In the Kinome database (see Figure \ref{fig:overlap}), there is an overlap between the information obtained from Kinase Protein pairs and Kinase Protein Phosphosite level triples. However, there is a discrepancy in the numbers. The overlap consists of 72 deleted kinases when considering the Kinase Protein pairs, but only 58 deleted kinases when extracting the information from the Kinase Protein Phosphosite level triples. This difference may be due to errors in formatting or the omission of a subset of data in one of the database files. Looking at the Kinase Interaction database as a source, the overlap depends on the cutoff used. Cutoffs of 6.73 and 4.72 correspond to a P value of less than 0.01 and 0.05, respectively, for strict and lenient lists of kinase-protein pairs as were suggested in the article. Accordingly the overlap between the deleted kinases and the two lists of Kinase and their target Protein pairs was 52 and 58 for the strict and lenient cutoffs, respectively. As will be demonstrated in the results section of this paper, the poor overlap was addressed by considering combinations of background knowledge sources. 

\clearpage

\section{Results}
\subsection{Silence-of-K Rule iterations and Kinase Compensation Rule}

Collecting the results was accomplished following a pipeline briefly described here. The Prolog interpreter was used to perform queries across all kinase deletions, and the resulting output was stored in lists. These lists were then parsed by Python scripts to perform additional visualisation and analysis in RStudio. This centered around the number of deletions that were predicted to have worked. Additionally, RStudio was used to create the kinase compensation networks, as described below in this section. \\

Using kinase-phosphosite level associations, the first iteration of the \textit{rule} proved to exhibit the best recall, confirming 21 kinase gene deletions having taken place, thus confirming the ground truth established by PCR, as reported in the study. Successive iterations of the \textit{rule}, 2nd (checking uniqueness) and 3rd (checking majority of known phosphosites) were less accurate, both confirming only 5 kinases as being silenced. Using kinase-protein level associations the results improved overall depending on the source of kinase protein pairs used. For the Kinome Database from bioGrid, 33, 7 and 3 kinase deletions were confirmed for \textit{rule} iterations 1, 2 and 3, respectively. Kinase Interaction Database (KID) with a lenient cut-off yielded 26, 9 and 5, whereas with a strict cut-off the results were 21, 8 and 4. Results summarised in Table 2. \\

As the overlap with a ground truth (the deletion of a kinase encoding gene) was poor, it was imperative to look at ways of improving this. Firstly, combining the outputs of the two levels of background knowledge considered was the most straightforward approach. Starting with the Kinome Database as a Kinase-Protein association  source, 54, 12 and 8 ground truths were captured with each successive iteration of the \textit{rule}. From the Kinase Interaction Database, the lenient cut off gave 47, 14 and 10, and with the strict cut off 42, 13 and 9. Overall we noted a significant improvement which was mainly evident in the first iteration of the rule, picking up increasingly more of the ground truth. These can be considered True Positive results. \\

\clearpage
\begin{table}[ht]
\centering
\begin{tabular}{|l|c|c|}
\hline
\begin{tabular}[c]{@{}l@{}}Rule Iteration \&\\ Background Knowledge\\ Level (Source where appropriate)\end{tabular} & True Positive & False Negative \\ \hline
First Iteration \& Phosphosite level & 21 & 63 \\ \hline
Second Iteration \& Phosphosite level & 5 & 79 \\ \hline
Third Iteration \& Phosphosite level & 5 & 79 \\ \hline
First Iteration \& Protein level (bioGrid) & 33 & 51 \\ \hline
Second Iteration \& Protein level (bioGrid) & 7 & 77 \\ \hline
Third Iteration \& Protein level (bioGrid) & 3 & 81 \\ \hline
First Iteration \& Protein level (KID Strict) & 21 & 63 \\ \hline
Second Iteration \& Protein level (KID Strict) & 8 & 76 \\ \hline
Third Iteration \& Protein level (KID Stict) & 4 & 80 \\ \hline
First Iteration \& Protein level (KID Lenient) & 26 & 68 \\ \hline
Second Iteration \& Protein level (KID Lenient) & 9 & 75 \\ \hline
Third Iteration \& Protein level (KID Lenient) & 5 & 89 \\ \hline
\end{tabular}
\caption[Rule iterations 1, 2 and 3 and corresponding TP and FN]{Table listing the True Positive and False Negative counts for each of the rule iterations. The False Negatives refer to kinase gene deletions we were not able to pick up due to either poor background information overlap or \textit{rule} iterations not performing as intended.}
\end{table}

\begin{figure}
    \centering
\includegraphics[width=.45\textwidth]{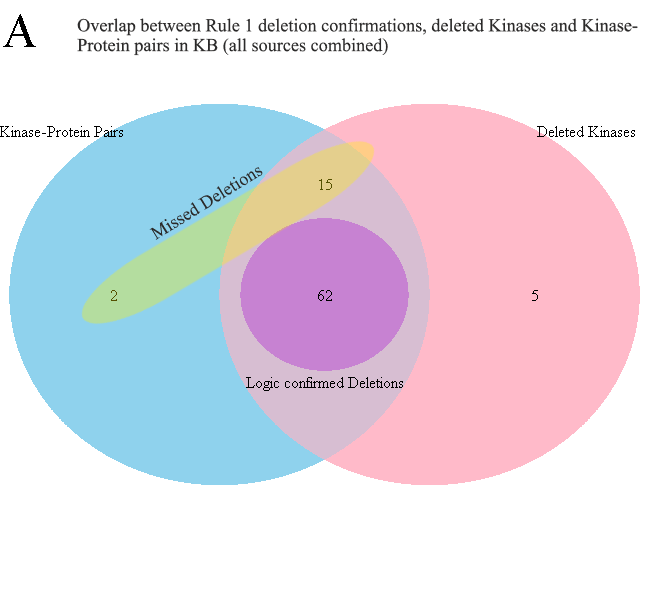}
\includegraphics[width=.45\textwidth]{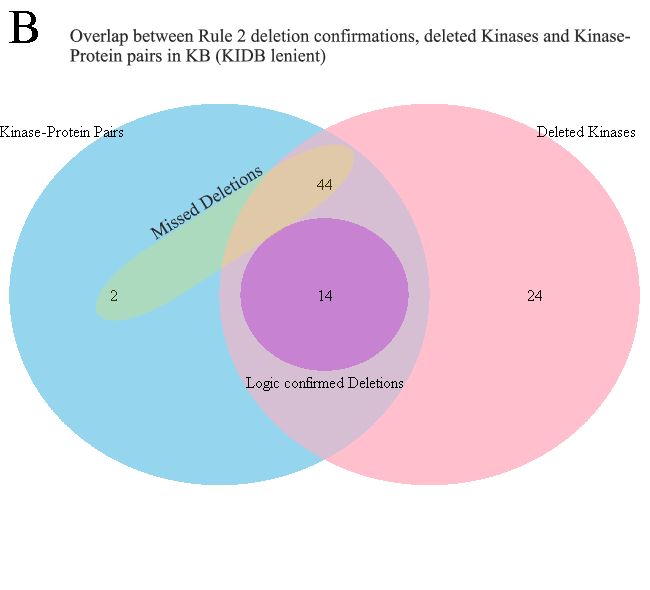}
\includegraphics[width=.5\textwidth]{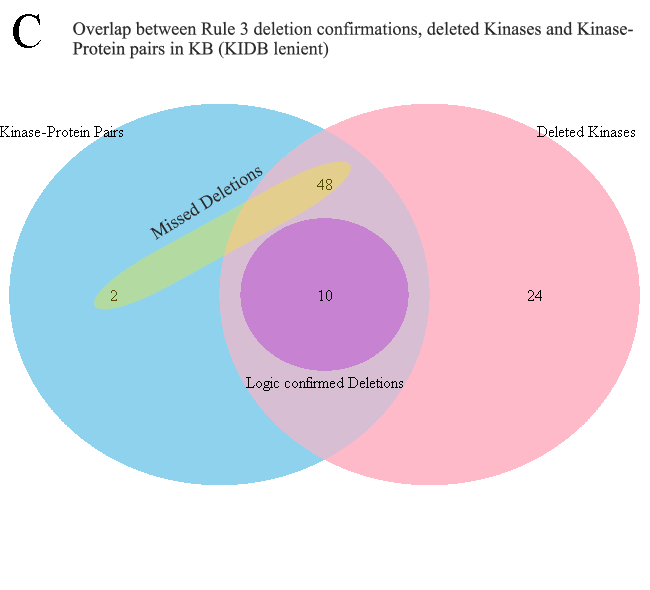}
\captionsetup{font=small}
    \caption[Best scoring combination results overlap] {Figure containing Venn Diagrams depicting the ground truth (true positives) and false negatives captured by the top performing \textit{rule} iterations and background knowledge levels considered. In all three Venn diagrams (A,B and C) the colour scheme is as follows: Purple for True Positives, light yellow for False Negatives, light blue for background knowledge kinase protein target pairs and light pink for gene deletion kinases. A: \textit{Rule} iteration 1 with background knowledge at the protein level and from all sources combined had the best recall with 62 True Positives identified out of a potential 78. B: Rule iteration 2, Kinase Interaction database lenient cut-off and phosphosite level information with 14 True Positives and 46 False Negatives out of 60 potential. C: The third iteration of the rule performed best when the background knowledge was sourced from KID with lenient cutoff at 10 True Positives, 50 False Negatives out of a potential 64. The difference in potential totals comes can be attributed to how many of the deleted kinases were present and therefore queriable in each instance of background knowledge source and protein or phosphosite level consideration. }
    \label{fig:overlap3}
\end{figure}

The next step was to combine all of the known background sources whilst removing duplicates to avoid clashes. This yielded a kinase-protein pair background knowledge base which covered (with at least one association) 79/82 of the deleted kinases in our dataset. This was the largest overall coverage and yielded the best result, confirming 62 individual deletions as seen in Figure \ref{fig:overlap3}. In contrast, iterations two and three yielded worse results, 13 and 8, respectively. From the above the best combinations between knowledge base sources (KID: strict/lenient and KINOME:bioGRID), level of background knowledge considered (kinase-phosphosite, kinase-protein or both) for each rule iteration were:

\begin{itemize}
    \item Rule 1: All BK, both levels: 62 TP / 17 FN
    \item Rule 2: KID Lenient cut-off, both levels: 14 TP / 46 FN
    \item Rule 3: KID Lenient cut-off, both levels: 10 TP / 50 FN
\end{itemize}

\clearpage

\subsubsection{Explaining the 17 missed deletions with a new rule}

From the above it was decided to further focus on explaining the results obtained from \textit{rule} iteration 1, with the results of both levels included and all background knowledge sources combined. However, as can be seen in the triple Venn diagram of Figure \ref{fig:overlap3} A, there were 17 kinase deletions the \textit{rule} iteration was not able to identify. As this iteration did not exclude shared targets between kinases a possible hypotheses is that other kinases are acting in a compensatory manner. For this the 'ykinasecompcheck/2' \textit{rule} was developed.

In brief, it checks whether a kinase (K2) increases its activity when a different kinase (DelK1) is silenced, in order to compensate for the loss of K1 activity and maintain the overall integrity of the signalling pathway. Specifically the  Prolog notation the \textit{rule} took the following form: \\

\begin{lstlisting}[style=prologStyle]
ykinasecompcheck(DelK1, K2) :-
    (perturbs(DelK1,_Pst,Tprot1,down,_) ;
    perturbs(DelK1,_Pst,Tprot1,unaffected,_)),
    sharedKinasetarget(DelK1,K2,Tprot1),
    knowntarget(K2, Pst1, Tprot1),
    (perturbs(DelK2,Pst1,Tprot1,up,_),
    (DelK2 \= K2, DelK2 \= DelK1)).
\end{lstlisting}

Specifically, the first goal 'perturbs/5' holds if the first argument (DelK1) perturbs the fourth argument (TProt1) in the direction specified by the fifth argument (down or unaffected). 'sharedKinasetarget/3' is a goal that is satisfied if the first argument (DelK1) and the second argument (K2) share the third argument (Tprot1) as a target. The information used for this, in order to maintain consistent background knowledge was the combination of all kinase protein background knowledge sources. 'knowntarget/3' is a goal that is satisfied if Tprot1 is a known target of the K2 at the phosphosite specified by the second argument (Pst1). Finally, (6) holds if the first argument (DelK2) perturbs the second argument (Pst1) in the direction specified by the fifth argument (up). The condition is true if DelK2 is different from both K2 and DelK1. The last goal ensures that kinase K2 indeed phosphorylates phosphosite Pst1 (in the condition it, itself is not deleted). 

Using the above rule and the list of 17 missed deleted Kinases, for each, a further list of compensatory kinases was collated. These lists where then used to build compensatory kinase interaction networks in order to visualise the effect of kinase deletions on known kinase interaction networks. Predicted or compensatory pathways were compared with pathways reconstructed from literature  \cite{Sharifpoor2011}. Visualisations were carried out in Rstudio using the igraph package.

\begin{figure}
    \centering
\includegraphics[width=.50\textwidth]{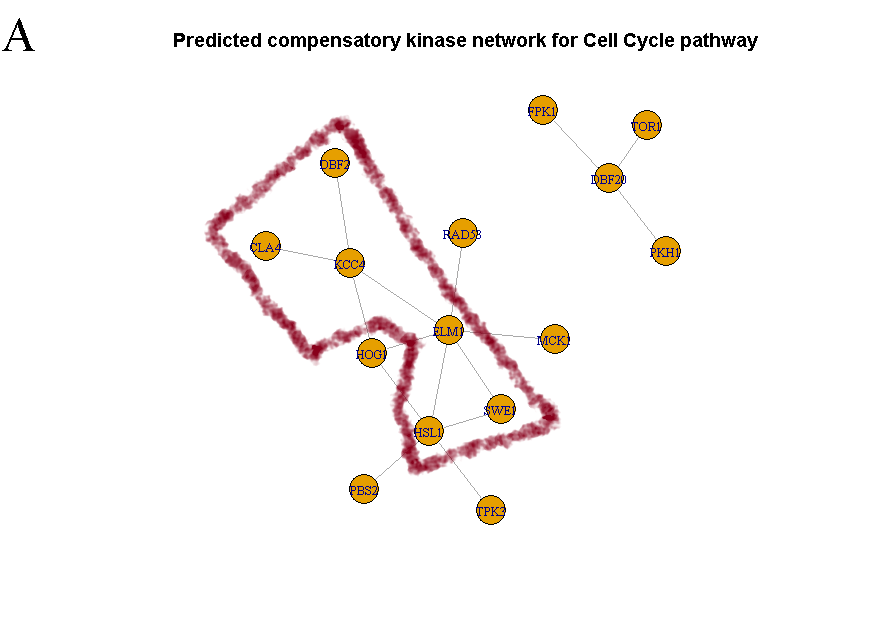}
\includegraphics[width=.50\textwidth]{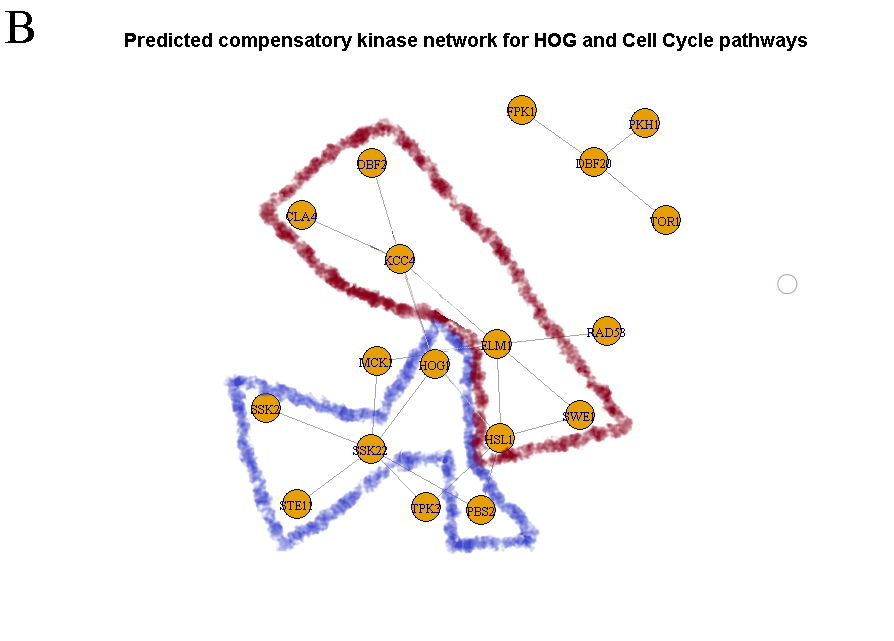}
\includegraphics[width=.55\textwidth]{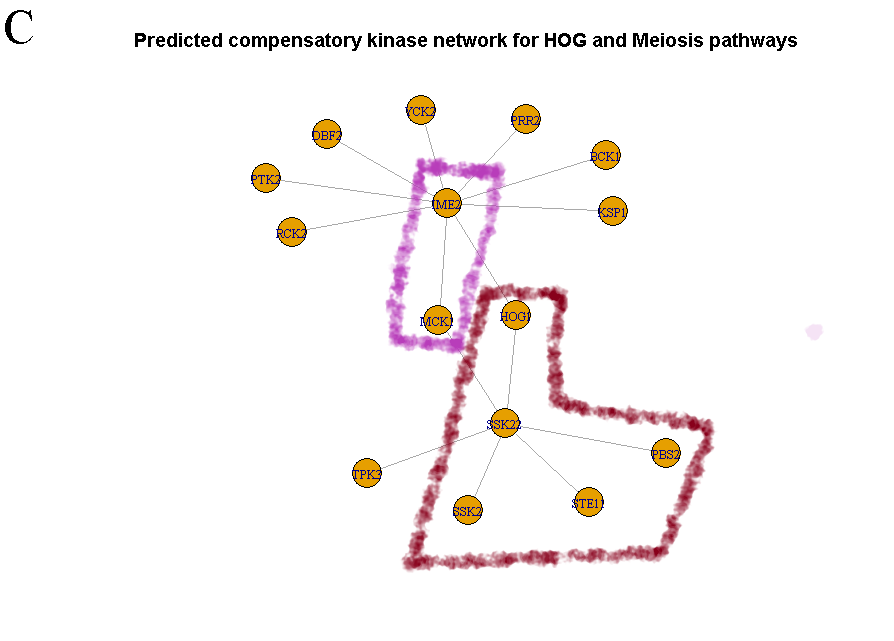}
\captionsetup{font=small}
    \caption[Predicted Kinase Compensatory Networks] {Predicted manner of kinase network rewiring following gene deletions. Kinases within pathways act in a compensatory manner. A) Cell Cycle pathway, predicted pathway rewiring. B) High-osmorality glucose and cell cycle pathway rewiring. C) High-osmorality and Meiosis pathway rewiring. Highlighted are the kinases belonging to the respective pathways that include False Negative i.e. missed gene deletions within our predictions.}
    \label{fig:compensatory predictions}
\end{figure}

None of the false negatives (kinase deletions not present in the \textit{rule} predictions) were present in the pathways associated with endocytosis, transcription response and mating. Cell cycle had the largest number of kinases compensating for ones from within our list. In Figure \ref{fig:compensatory predictions} Cell Cycle pathway is highlighted with a red border, the High-Osmorality glucose (HOG) pathway with a blue border and the meiosis with a fuchsia border. They are all connected canonically, showing the relative accuracy of the \textit{rule} described above in picking up the shift in pathway activity to account for the deletion of a given kinase within the pathway. For the highlighted pathways the majority of their constituent kinases are present, deletions of which the first iteration of the \textit{rule} did not pick up. This indirectly suggests that in response to certain key deletions these highlighted pathways are able to rewire themselves in an effort to counteract the perturbation. 

\clearpage

\section{Discussion}
Despite \textit{Saccharomyces cerevisiae} being extensively studied as a model system, there was a notable lack of coverage between kinase-phosphosite associations and the observational (phosphoproteomics) data obtained from the gene knockout study. Therefore implementing the methodology developed initially on the Toy Model, proved to be more challenging than anticipated.

The approach that yielded the most accurate results, in terms of recall, with the simplest iteration of the \textit{rule}. The issue of poor overlap was also partially overcome by combining background knowledge sources and levels of granularity. Increasing complexity of \textit{rules} did not yield better results.

Delving deeper into the False Negatives, i.e. kinases from verified deleted genes which we were unable to pick up, yielded interesting results. The interconnected nature of kinase interaction pathways was captured by the 'ykinasecompcheck/2'  \textit{rule}. Using the 17 kinases as a base we were able to identify and contextualise the majority of major large Yeast pathways, namely HOG and cell cycle as well as smaller one such as Meiosis. For the remaining pathways, their constituents were correctly captured as having been silenced. Combining the outputs of the individual silenced gene queries offers an insight into potential yeast signalling rewiring following strong perturbation such as silencing of a group of kinases.

The issue with overlap can be addressed by initiatives such as bioGRID and other datasets providing consistent information at both the protein and phosphosite level interactions. This problem is even greater and more present in the human kinome and proteome.  However, combining resources, a crude approach at best, seemed to alleviate this, at least partially. A recently published study \cite{Johnson2023} made a significant contribution in assigning substrates to a large number of active Ser/Thr kinases (approximately 84$\%$) in humans.

Next steps in applying this methodology would involve augmenting \textit{facts} with probabilities, via the ProbLog2 \cite{probLog2} framework. Specifically, the immediate aim of this would be to capture the epistemic uncertainty in the kinase phosphosite associations as well as the mix of alleatoric and epistemic uncertainty that characterises the process of mass spectrometry based data acquisition and related post processing. The resulting output could further increase confidence in the predictions. Both of these require the mining of appropriate and well curated information in order to prove viable.

\nocite{*}
\bibliographystyle{eptcs}
\bibliography{zbibliography}
\end{document}